\documentclass[aoas,preprint]{imsart}

\RequirePackage[OT1]{fontenc}
\RequirePackage{amsthm,amsmath}
\RequirePackage{natbib}
\RequirePackage[colorlinks,citecolor=blue,urlcolor=blue]{hyperref}

\usepackage{afterpage,rotating,latexsym,psfrag,epsf,epsfig,amssymb,subfigure,color}
\usepackage[mathlines]{lineno}

\startlocaldefs
\numberwithin{equation}{section}
\theoremstyle{plain}

\endlocaldefs
\newcommand{\bfu}{\hbox{\boldmath$u$}}
\newcommand{\bfn}{\hbox{\boldmath$n$}}
\newcommand{\bftheta}{\hbox{\boldmath$\theta$}}

\newcommand{\bfx}{\hbox{\boldmath$x$}}

\newcommand{\bfeta}{\hbox{\boldmath$\eta$}}
\newcommand{\bfeps}{\hbox{\boldmath$\epsilon$}}

\newcommand{\bfbeta}{\hbox{\boldmath$\beta$}}
\newcommand{\bfepsilon}{\hbox{\boldmath$\epsilon$}}

\newcommand{\bfzero}{\hbox{\boldmath$0$}}

\newcommand{\bfomega}{\hbox{\boldmath$\omega$}}
\newcommand{\bfsomega}{\hbox{\boldmath{\footnotesize{$\omega$}}}}
\newcommand{\bftomega}{\hbox{\boldmath{\tiny{$\omega$}}}}

\newcommand{\comment}[1]{}

\begin{document}


\begin{frontmatter}

\title{Capture-recapture abundance estimation using a semi-complete data likelihood approach}
\runtitle{Semi-complete data likelihood}

\begin{aug}
\author{\fnms{Ruth} \snm{King}\thanksref{m1}\ead[label=e1]{Ruth.King@ed.ac.uk}},
\author{\fnms{Brett T.} \snm{McClintock}\thanksref{m2}\ead[label=e2]{Brett.McClintock@noaa.gov}},
\author{\fnms{Darren} \snm{Kidney}\thanksref{m3}\ead[label=e3]{darrenkidney@googlemail.com}}
\and
\author{\fnms{David} \snm{Borchers}\thanksref{m3}\ead[label=e4]{dlb@st-andrews.ac.uk}}

\runauthor{R. King, B. McClintock, D. Kidney and D. Borchers}

\affiliation{University of Edinburgh\thanksmark{m1}, NOAA National Marine Mammal Laboratory\thanksmark{m2} and University of St Andrews\thanksmark{m3}}

\address{Address of King\\
School of Mathematics, University of Edinburgh, \\
James Clerk Maxwell Building, \\
The King's Buildings, Peter Guthrie Tait Road, \\
Edinburgh, UK. EH9 3FD. }

\address{Address of McClintock\\
National Marine Mammal Laboratory, \\
Alaska Fisheries Science Center, \\
NOAA National Marine Fisheries Service,  \\
7600 Sand Point Way NE, Seattle, \\
Washington 98115 USA}

\address{Address of Kidney and Borchers\\
Centre for Research into Ecological and Environmental Modelling \\
and School of Mathematics and Statistics, \\ The Observatory, Buchanan Gardens, \\
University of St Andrews, St Andrews, UK. KY16 9LZ. }

\printead{e1}

\end{aug}

\label{firstpage}

\begin{abstract}
\comment{\begin{itemize}
\item Present efficient model-fitting approach in the estimation of closed populations using capture-recapture data in the presence of individual heterogeneity.
\item Split the likelihood into two component parts: a complete data likelihood conditioning on (unobserved) individual effects (missing covariate values and/or random effects) for individuals observed within the study and a data augmentation approach; and an observed data likelihood for the individuals not observed within the study approximated using a numerical integration approach.
\item Combine advantages of the two different approaches (data augmentation and numerical integration) for component parts - only one integral is needed for all unobserved individuals; whereas data augmentation can be efficient for observed individuals.
\item This approach permits the model-fitting in BUGS (as for data augmentation) but is more flexible with regard to prior specification on the total population size and removal of the super-population approach when using data augmentation that by necessity specifies an upper bound on the population and can be computationally inefficient when a large super-population needs to be specified.
\item Apply the approach to a closed capture-recapture population model $M_h$ and spatially explicit capture-recapture (SECR) model.
\item Compare the performance of the different algorithms using real datasets. 
\end{itemize} }

Capture-recapture data are often collected when abundance estimation is of interest. In this manuscript we focus on abundance estimation of closed populations. In the presence of unobserved individual heterogeneity, specified on a continuous scale for the capture probabilities, the likelihood is not generally available in closed form, but expressible only as an analytically intractable integral. Model-fitting algorithms to estimate abundance most notably include a numerical approximation for the likelihood or use of a Bayesian data augmentation technique considering the complete data likelihood. We consider a Bayesian hybrid approach, defining a ``semi-complete'' data likelihood, composed of the product of a complete data likelihood component for individuals seen at least once within the study and a marginal data likelihood component for the individuals not seen within the study, approximated using numerical integration. This approach combines the advantages of the two different approaches, with the semi-complete likelihood component specified as a single integral (over the dimension of the individual heterogeneity component). In addition, the models can be fitted within BUGS/JAGS (commonly used for the Bayesian complete data likelihood approach) but with significantly improved computational efficiency compared to the commonly used super-population data augmentation approaches (between about 10 and 77 times more efficient in the two examples we consider). The semi-complete likelihood approach is flexible and applicable to a range of models, including spatially explicit capture-recapture models. The model-fitting approach is applied to two different datasets: the first relates to snowshoe hares where model $M_h$ is applied and the second to gibbons where a spatially explicit capture-recapture model is applied. 

\end{abstract}


\end{frontmatter}

\section{Introduction}

\comment{
\begin{itemize}
\item Closed population models with individual effects to date have typically used either a data augmentation or trans-dimensional approach using a complete data likelihood (or auxiliary variable approach); or numerical integration using the observed data likelihood. Note - also Bonner and Schofield approach - using MC(MC)MC approach.
\item The data augmentation approach can be fitted in BUGS, but requires the specification of a super-population. 
\item The super-population essentially provides an upper bound for the total population and is directly related to computational speed.
\item In addition, the data augmentation approach (as implemented in BUGS) specifies an implicit prior on the total population (discrete uniform for U[0,1] prior on $\psi$; truncated at the upper limit) - see Schofield and Barker (2014) and Link (2013) papers.
\item Aim is to provide an alternative model fitting approach, using a numerical integration approach instead of a data augmentation approach - more flexible with regard to prior specification on $N$; incorporation of covariates - still fittable in WinBUGS. 
\end{itemize}}

In order to estimate total abundance capture-recapture data are often collected on the population under study. Capture-recapture data collection methods involve partially observing the population at a series of capture events (or using a number of different sources), such that each individual observed within the study is uniquely identifiable. Assuming that marks are unique and cannot be lost, a capture history for each individual observed within the study can be constructed, detailing whether the given individual is observed or not at each capture event. Statistical models can be constructed and applied to capture-recapture data to estimate the number of individuals in the population that are not observed. We focus on closed population models, where it is assumed that that there are no births/deaths/migrations in the population within the study period. Applications include estimating the number of injecting drug users \citep{kinbohh14,ovekbhh14}, pages on the world wide web \citep{fie99}, disease prevalence \citep{manf08} and animal populations \citep{bor02}. We focus on statistical models for ecological data where individuals are observed at a series of capture events. For further discussion of ecological (closed) capture-recapture data, and the underlying assumptions, see for example, \citet{bor02}, \citet{wilnc02} and \citet{mccm14}. 

In general, the likelihood of capture-recapture data can be expressed in multinomial form, where the different multinomial cells correspond to each possible capture history and the cell entries to the number of individuals with the given capture history. The unknown parameters to be estimated in the likelihood function are the capture (or detection) probabilities and the total population size (or number of individuals in the population unobserved at any capture event). \cite{otis78} described three different possible effects on the capture probabilities corresponding to temporal ($t$), behavioural ($b$) and individual heterogeneity ($h$) effects. We adopt the standard notation and describe the different models by $M_{a}$, such that $a \subseteq \{t,b,h\}$, corresponding to the combination of effects in the given model. 

In this paper we focus on models that include individual heterogeneity (i.e. $M_h$-type models). Individual heterogeneity is often introduced by specifying the capture probabilities as a finite or infinite mixture. Finite mixture models lead to an explicit likelihood expression which can be maximised numerically to obtain the maximum likelihood estimates (MLEs) of the parameters of interest \citep{ple00}. Infinite mixture models specify the individual heterogeneity as a random effects model. For the special case of a Beta-Binomial random effects component the likelihood is available in closed form \citep{dorr03,morr08}. We will consider the more general case, with an arbitrary individual heterogeneity component leading to an analytically intractable likelihood. Previous approaches to fit such models to the data include (i) numeral integration to estimate the \emph{marginal} (or observed) data likelihood \citep{coua99,bore08,gimc10}; and (ii) Bayesian data augmentation techniques, using a \emph{complete} data likelihood approach (corresponding to the joint probability density function of the capture histories and individual effects), integrating out the individual heterogeneity component within a Markov chain Monte Carlo-type (MCMC) algorithm \citep{dure05,roydl07,roykgk09,kinb08,kinmgb09,royd12}. We combine these two approaches defining a \emph{semi-complete} data likelihood constructed as the product of a complete data likelihood component for the individuals seen at least once in the study and a marginal data likelihood component for the unseen individuals. This combines the advantages of each of the individual approaches. We note that similar approaches have been previously proposed for specific applications, using bespoke computer codes. Most notably, \cite{fie99} propose a conditional MCMC algorithm for Rasch-type models, employing a block update of the total population size and individual heterogeneity terms; while \cite{bons14} consider a similar Monte Carlo in MCMC approach applied to individual covariate models. We describe how the latter approach is a special case of our general semi-complete data likelihood approach in Section \ref{special}. Finally, we demonstrate how individual heterogeneity models can be efficiently fitted using BUGS/JAGS with general prior structures specified on all the model parameters (including the total population size) and provide the associated computer codes. 

The paper proceeds as follows. Section \ref{sect2} describes the general closed population model structure and associated notation. Section \ref{sect3} describes previous model-fitting approaches and the new proposed semi-complete data likelihood approach. The implications of the BUGS/JAGS specification for the semi-complete data likelihood and previous Bayesian complete data likelihood approaches are compared in Section \ref{sect4} and the approaches applied and compared for two real examples: the first example relates to snowshoe hares where model $M_h$ is applied and the second to a dataset of gibbons where a spatially explicit capture-recapture model is applied. Finally in Section \ref{sect5} we conclude with a discussion. 

\section{Individual heterogeneity models}\label{sect2}

We assume that within the capture-recapture study there is a series of $T$ discrete capture occasions. Within the study a total of $n$ distinct individuals are observed, with the total (unknown) population size denoted by $N$. For simplicity we arbitrarily number the observed individuals $i=1,\dots,n$ and the unobserved individuals $i=n+1,\dots,N$. Let $p_{it}$ denote the capture probability of individual $i=1,\dots,N$ at time $t=1,\dots,T$. Further, for standard capture-recapture data, $\bfx_i = \{x_{it}: t=1,\dots,T\}$ denotes the capture history of individual $i=1,\dots,N$, such that
\begin{linenomath}
\[
x_{it} = \left\{\begin{array}{cl} 0 & \mbox{ individual $i$ is unobserved on occasion $t$;} \\
1 & \mbox{ individual $i$ is observed on occasion $t$.} \end{array}\right.
\]
\end{linenomath}
\comment{We note that the observed data $\bfx = \{\bfx_1,\dots,\bfx_n\}$ is often augmented with additional observed individual covariate information. For example, for SECR the covariate information is time-dependent and corresponds to trap location(s). Alternatively, time-invariant and/or time-varying discrete/continuous individual covariates may be recorded when an individual is observed. }

We consider individual heterogeneity specified such that
\begin{linenomath}
\[
p_{it} = g(\bftheta, \bfeps_i),
\]
\end{linenomath}
for some function $g$, where $\bftheta$ denotes the model parameters associated with the capture probabilities (which may include, for example, temporal and/or behavioural effect terms, regression coefficients for covariate values etc.) and $\bfepsilon = \{\bfeps_i:i=1,\dots,N\}$ such that $\bfeps_i \in \mathbb{S} \subset \mathbb{R}^k$, corresponding to the individual heterogeneity term for individual $i=1,\dots,N$. Further, we assume an underlying model for the individual heterogeneity, such that $\bfepsilon$ is a function of the parameters $\bfeta$, and that the individual heterogeneity terms, $\bfepsilon_i$, are independent of each other conditional on $\bfeta$.  The associated joint probability density function of the heterogeneity terms is given by $f_{\bfepsilon}(\bfepsilon | N, \bfeta) = \prod_{i=1}^N f_{\epsilon}(\bfepsilon_i | \bfeta)$, using the conditional independence assumption (and dropping the dependence on $N$ for the conditional density function of the individual heterogeneity terms for individual $i$). Further, to provide a general framework for both observed and unobserved individual heterogeneity we additionally write $\bfeps = \{\bfeps^{Obs},\bfeps^{Mis}\}$ where $\bfeps^{Obs}$ denotes the set of observed individual heterogeneity components and $\bfeps^{Mis}$ the set of unobserved individual heterogeneity components. Similarly, we write  $\bfeps_i = \{\bfeps_i^{Obs}, \bfeps_i^{Mis}\}$, for $i=1,\dots,N$ with obvious notation. Finally, we assume that the capture histories of the individuals are independent of each other given the capture probability model parameters, $\bftheta$, and individual heterogeneity terms, $\bfepsilon$. 

The \emph{marginal} data likelihood can be expressed in the form,
\begin{eqnarray}
f_m(\bfx, \bfeps^{Obs}| N, \bftheta, \bfeta) & = & \int_{\bfepsilon_1^{Mis}}\dots\int_{\bfepsilon_N^{Mis}} f_c(\bfx, \bfepsilon | N, \bftheta, \bfeta) d\bfepsilon_1^{Mis} \dots d\bfepsilon_N^{Mis} \nonumber \\
& = & \int_{\bfepsilon_1^{Mis}}\dots\int_{\bfepsilon_N^{Mis}} f_{\bfx}(\bfx| N, \bftheta,\bfepsilon) f_{\bfepsilon}(\bfepsilon | N, \bfeta) d\bfepsilon_1^{Mis} \dots d\bfepsilon_N^{Mis}
\nonumber \\
& \propto & \frac{N!}{(N-n)!} \prod_{i=1}^N \int_{\bfepsilon_i ^{Mis}} f_x(\bfx_i| \bftheta,\bfepsilon_i) f_{\epsilon}(\bfepsilon_i | \bfeta) d\bfepsilon_i^{Mis},
\label{lik:obs}
\end{eqnarray}
using the multinomial distributional form of the capture-recapture data (omitting the constant multinomial coefficients for simplicity), and conditional independence of the random effect terms. The term $f_c(\bfx, \bfepsilon | N, \bftheta, \bfeta)$ corresponds to the \emph{complete} data likelihood (i.e. the joint probability density function of the capture histories and individual effects); $f_{\bfx}(\bfx| N, \bftheta,\bfepsilon)$ the conditional likelihood of the capture histories (where the conditioning includes the individual heterogeneity terms); and $f_{\bfepsilon}(\bfepsilon | N, \bfeta)$ the joint probability density function of the individual heterogeneity terms. The term $f_x(\bfx_i| \bftheta,\bfepsilon_i)$ corresponds to the conditional likelihood of capture history for individual $i=1,\dots,N$; and 
$f_\epsilon(\bfepsilon_i | \bfeta)$ the conditional probability density function of the individual heterogeneity component for individual $i=1,\dots,N$ (where in each case we drop the dependence on $N$). 

\subsection*{Example 1 - Continuous individual covariates}

We consider the case with $q$ time-invariant continuous individual covariates $\bfepsilon=\{\bfepsilon_1,\dots,\bfepsilon_N\}$ where $\bfepsilon_i \in \mathbb{S} \subseteq \mathbb{R}^q$ denotes the covariate values associated with individual $i=1,\dots,N$. Since the covariate values are time-invariant, the associated capture probabilities for each individual are also time-invariant, so that $p_{it} = p_i$ for $t=1,\dots,T$. Assuming that the capture probabilities are linearly related to the covariate values via some link function, we may specify,
\[
g^{-1}(p_{i}) = \alpha + \bfbeta^T \bfepsilon_i,
\]
so that $\bftheta = \{\alpha, \bfbeta\}$. Common choices for $g^{-1}$ include the logit and probit functions. Additional individual/temporal random effects can be included in the capture probabilities, but we omit these here for simplicity (see Example 2). Further we specify a parametric model for the covariate values, assuming that conditional on the additional covariate parameters $\bfeta$, the covariate values are independent.

Assuming that for each individual observed within the study the set of individual covariate values is recorded, we have that $\bfepsilon^{Obs} = \{\bfepsilon_i:i=1,\dots,n\}$ and $\bfepsilon^{Mis} = \{\bfepsilon_i:i=n+1,\dots,N\}$. More generally, the covariate values may not be recorded for all observed individuals. For example, the observation process may include sightings recorded from a distance (rather than physical captures) so that the covariate may not be able to be obtained if a physical capture is necessary (for example if the covariate corresponds to wingspan). In this case the set of unobserved individual heterogeneity terms is extended to include the  unknown covariate values for observed individuals. 

The complete data likelihood is of the form,
\begin{eqnarray*}
f_c(\bfx, \bfepsilon|N,\bftheta, \bfeta) & \propto & \frac{N!}{(N-n)!} \prod_{i=1}^N \left[\prod_{t=1}^T p_{it}^{x_{it}} (1-p_{it})^{1-x_{it}}\right] \times f_{\epsilon}(\bfepsilon_i| \bfeta) \\
& = & \frac{N!}{(N-n)!} \prod_{i=1}^N p_{i}^{y_i} (1-p_{i})^{T-y_{i}} \times f_{\epsilon}(\bfepsilon_i| \bfeta),
\end{eqnarray*}
where $p_{i}$ is of the above form and $y_i = \sum_{t=1}^T x_{it}$ (denoting the total number of times individual $i$ is observed). The first term of the complete data likelihood corresponds to the conditional likelihood (conditional on the individual covariate terms) and the second term to the individual covariate component. 

The marginal data likelihood integrates out the unobserved covariate values $\bfepsilon^{Mis}$. For notational simplicity we provide the marginal data likelihood for the special case where all covariate values are known for individuals observed within the study (i.e. $\bfepsilon^{Obs} = \{\bfepsilon_i:i=1,\dots,n\}$ and $\bfepsilon^{Mis} = \{\bfepsilon_i:i=n+1,\dots,N\}$):
\begin{eqnarray*}
f_m(\bfx, \bfepsilon^{Obs}| N, \bftheta,\bfeta) & \propto & \frac{N!}{(N-n)!} \prod_{i=1}^n p_{i}^{y_i} (1-p_{i})^{T-y_{i}} f_{\epsilon}(\bfepsilon_i| \bfeta) \times \prod_{i=n+1}^N \int_{\bfepsilon_i} p_{i}^{y_i} (1-p_{i})^{T-y_{i}} f_{\bfepsilon}(\bfepsilon_i| \bfeta) d \bfepsilon_i \\
& = & \frac{N!}{(N-n)!} \prod_{i=1}^n p_{i}^{y_i} (1-p_{i})^{T-y_{i}} f_{\epsilon}(\bfepsilon_i| \bfeta) \times \left[ \int_{\bfepsilon_0} (1-p_0)^T f_\epsilon(\bfepsilon_0|\bfeta) d \bfepsilon_0\right]^{N-n},
\end{eqnarray*}
where $g^{-1}(p_0) = \alpha + \bfbeta^T \bfepsilon_0$. The extension to the case where observed individuals may also have unknown covariate values is immediate. 

We note, in general, the model can be extended to include time-varying individual covariates, using the time and individual dependent capture probability, $p_{it}$. This typically substantially increases the number of unobserved covariate values, since if an individual is not observed, the corresponding covariate value is necessarily also unknown. However, for closed populations, to satisfy the condition that the population is closed the study period is generally short in duration so that changes in time-varying individual covariate values is likely to be limited. 

\subsection*{Example 2 - $M_h$-type models}

For $M_h$-type models the individual heterogeneity corresponds to an unobserved individual random effect component (so that $\bfepsilon^{Obs} = \emptyset$ and $\bfepsilon^{Mis} = \bfepsilon$). For example, for model $M_h$ we may set $\bftheta = \{\alpha\}$ and $\bfeta = \{\sigma^2\}$ such that,
\begin{linenomath}
\[
\epsilon_i \sim N(0,\sigma^2),
\]
\end{linenomath}
for $i=1,\dots,N$, where $\sigma^2$ denotes the individual random effect variance and $\mathbb{S} = \mathbb{R}$. For this model, the capture probabilities are again independent of time $t$, so we can write $p_{it} = p_i$ for all $t = 1,\dots,T$, with
\begin{linenomath}
\[
g^{-1}(p_i) = \alpha + \epsilon_i,
\]
\end{linenomath}
for $i=1,\dots,N$ and $t=1,\dots,T$. Common choices for $g^{-1}$ include the logit and probit functions. The extension to incorporate additional time and/or behavioural effects is immediate \citep[i.e. models $M_{th}, M_{bh}$ and $M_{tbh}$; see for example][]{kinb08}. 

The complete data likelihood for model $M_h$ can be written in the form,
\begin{eqnarray*}
f_c(\bfx, \bfepsilon|N,\bftheta,\bfeta) & \propto & \comment{\frac{N!}{n!(N-n)!} \prod_{i=1}^N \prod_{t=1}^T p_{i}^{x_{it}} (1-p_{i})^{1-x_{it}} \times \prod_{i=1}^N \frac{1}{\sqrt{2 \pi \sigma^2}} \exp\left(-\frac{\epsilon_i^2}{2 \sigma^2}\right) \\
& = &} \frac{N!}{(N-n)!} \prod_{i=1}^N p_{i}^{y_i} (1-p_{i})^{T-y_{i}} \times \frac{1}{\sqrt{2 \pi \sigma^2}} \exp\left(-\frac{\epsilon_i^2}{2 \sigma^2}\right),
\end{eqnarray*}
where $p_{i}$ is of the above form and $y_i = \sum_{t=1}^T x_{it}$. Once again, the first term of the complete data likelihood corresponds to the conditional likelihood (conditional on the individual random effect terms) and the second term to the individual effect component. 

The marginal data likelihood integrates out the $\bfepsilon$ terms and (dropping the term $\bfepsilon^{Obs}$ since no individual heterogeneity terms are observed, i.e. $\bfepsilon^{Obs} = \emptyset$) can be efficiently expressed as,
\begin{eqnarray*}
f_m(\bfx| N, \bftheta,\bfeta) & \propto & \frac{N!}{(N-n)!} \prod_{k=0}^T \left[ \int_{\epsilon_k \in \mathbb{S}} (p_k)^{k} (1-p_k)^{T-k} \frac{1}{\sqrt{2 \pi \sigma^2}} \exp\left(-\frac{\epsilon_k^2}{2 \sigma^2}\right) d\epsilon_k \right]^{n_k} ,
\end{eqnarray*}
where $n_k = \sum_{i=1}^N I(y_i = k)$ and denotes the number of individuals observed $k$ times within the study, for $k=0,\dots,T$ (so that $n_0$ is unobserved and $N = n_0 + n$) and $g^{-1}(p_k) = \alpha + \epsilon_k$.

\comment{This simplifies under model $M_h$ since $p_{it} = p_i$ for all $t=1,\dots,T$. Let $f_k$ denote the number of individuals observed a total of $k$ times within the study, for $k=0,\dots,T$ (so that $f_0$ is unobserved and $N = f_0 + n$). Then, we can simplify the marginal data likelihood such that,
\[
f(\bfx| N, \bftheta) = \frac{N!}{n!(N-n)!} \prod_{k=0}^T \left[ \int_{\epsilon_k \in \mathbb{S}} (p_k)^{k} (1-p_k)^{T-k} \frac{1}{\sqrt{2 \pi \sigma^2}} \exp\left(-\frac{\epsilon_k^2}{2 \sigma^2}\right) d\epsilon \right]^{f_k} ,
\]
with $g^{-1}(p_k) = \alpha + \epsilon_k$.}

\subsection*{Example 3 - SECR models}

For traditional spatially explicit capture-recapture models, $\mathbb{S} \subset \mathbb{R}^2$ and the individual heterogeneity corresponds to the unobserved activity centre of the individual (so that $\bfepsilon^{Obs} = \emptyset$ and $\bfepsilon^{Mis} = \bfepsilon$). The range of possible models is greater for SECR than non-spatial capture-recapture as SECR models involve multiple traps or detectors at different locations on each occasion and take account of the location(s) of observations within occasions. To this end we define $\bfu_j=(u_{j1},u_{j2})\in\mathbb{R}^2$ to be the Cartesian coordinates of trap $j$, for $j=1,\ldots,J$. We consider the likelihood for a study with binary detection data within occasion, such that
\begin{linenomath}
\[
x_{ijt} = \left\{\begin{array}{cl} 0 & \mbox{ individual $i$ is unobserved by detector $j$ on occasion $t$;} \\
1 & \mbox{ individual $i$ is observed by detector $j$ on occasion $t$.} \end{array}\right.
\]
\end{linenomath}
We consider the case where individuals can be observed by more than one detector at each occasion and we assume that observations by different detectors within occasions (as well as between occasions) are independent. In this context, $\bfepsilon_i=(\epsilon_{i1},\epsilon_{i2})\in \mathbb{R}^2$ ($i=1,\ldots,N$) denote the Cartesian coordinates of the activity centres of the $N$ individuals in $\mathbb{S}\subset\mathbb{R}^2$. It is usually assumed that these are independently uniformly distributed in $\mathbb{S}$ and do not change between occasions, so that $f_{\bfepsilon}(\bfepsilon|N,\bfeta)=\prod_{i=1}^N f_{\epsilon}(\bfepsilon_i|\bfeta)=A^{-N}$, where $A$ is the area of $\mathbb{S}$. The probability of individual $i$ being observed by detector $j$ at capture occasion $t$, denoted $p_{ijt}$ is assumed to depend on only the distance of the detector from the activity centre of individual $i$, so that $p_{ijt} = g(\bftheta, ||\bfu_j-\bfeps_i||)$, where $||\bfu_j-\bfeps_i||$ is the vector norm $\sqrt{\sum_{k=1}^2(u_{jk}-\epsilon_{ik})^2}$. The half-normal form is a common choice for $g$. For example, assuming that the capture probabilities are time-independent, we may specify,
\begin{linenomath}
\[
p_{ijt} = p_{ij} = p_0\exp\left(-\frac{||\bfu_j-\bfeps_i||^2}{2 \sigma^2}\right)
\]
\end{linenomath}
\noindent
with $\bftheta = \{p_0,\sigma^2\}$. 

The complete data likelihood can be written as
\begin{eqnarray*}
f_c(\bfx, \bfepsilon|N,\bftheta, \bfeta) & \propto & \frac{N!}{(N-n)!} \prod_{i=1}^N \left[ \prod_{t=1}^T \prod_{j=1}^J p_{ijt}^{x_{ijt}} (1-p_{ijt})^{1-x_{ijt}} \times  f_{\epsilon}(\bfepsilon_i|\bfeta)\right], 
\end{eqnarray*}
where $p_{ijt}$ is of the above form. The first term in the product over individuals corresponds to the conditional likelihood associated with individual $i$ (conditional on the individual random effect terms) and the second term to the corresponding individual effect component. 

The marginal data likelihood integrates out the $\bfepsilon_i$ terms and can be expressed as,
\begin{linenomath}
\[
f_m(\bfx | N, \bftheta, \bfeta)\propto \frac{N!}{(N-n)} \prod_{i=1}^N \int_{\bfepsilon_i^{Mis}} \prod_{t=1}^T\prod_{j=1}^J p_{ijt}^{x_{ijt}} (1-p_{ijt})^{1-x_{ijt}}  \times f_{\epsilon}(\bfepsilon_i|\bfeta) d\bfepsilon_i^{Mis},
\]
\end{linenomath}
once more omitting the term $\bfepsilon^{Mis} = \emptyset$. 

\subsection{Model fitting}

In the presence of individual heterogeneity leading to an analytically intractable marginal data likelihood a range of different approaches have been proposed. These include a (classical) numerical integration approach, approximating the marginal data likelihood and a (Bayesian) data augmentation approach using the complete data likelihood. For the particular application to $M_h$-type models and SECR, see for example \cite{coua99,bore08,gimc10} (for a classical numerical integration approach) and \cite{dure05,roydl07,roykgk09,kinb08,royd12} (for Bayesian data augmentation approaches). We briefly describe the approaches in turn. 

\subsubsection{Marginal data likelihood}\label{obs}

For a general individual heterogeneity model, the marginal data likelihood may not be available in closed form (exceptions exist where the heterogeneity component is described as a finite mixture model or infinite Beta distribution). In this case, the corresponding likelihood is given in equation (\ref{lik:obs}) as a product of integrals. For computational efficiency, we are able to combine like terms in the likelihood corresponding to each unique encounter history (corresponding to the combined capture history and observed individual heterogeneity values). Notationally, let $\Omega$ denote the set of possible encounter histories; $\bfx_{\bfsomega}$ the capture history for $\bfomega \in \Omega$; $\bfepsilon_{\bfsomega}$ the individual heterogeneity terms for encounter history $\bfomega \in \Omega$; $\bfepsilon_{\bfsomega}^{Mis}$ the unobserved individual heterogeneity terms for encounter history $\bfomega \in \Omega$ and $n_{\bfsomega}$ the number of individuals with encounter history $\bfomega$. The marginal data likelihood can be expressed as,
\begin{linenomath}
\[
f_m(\bfx,\bfepsilon^{Obs}|N,\bftheta,\bfeta) \propto \frac{N!}{(N-n)!} \prod_{\bfomega \in \Omega} \left[\int_{\bfepsilon_{\bftomega}^{Mis}} f_{\bfx}(\bfx_{\bfsomega} | N, \bftheta,\bfepsilon_{\bfsomega}) f_{\bfepsilon}(\bfepsilon_{\bfsomega} | N, \bfeta) d\bfepsilon_{\bfsomega}^{Mis} \right]^{n_{\bfsomega}}.
\]
\end{linenomath}
Thus, this likelihood requires the estimation of a series of integrals each of dimension (at most) dim($\mathbb{S}$), where typically dim($\mathbb{S}$) is small. For example, in the presence of $q$ time invariant continuous covariates, dim($\mathbb{S}) = q$, for model $M_h$, dim($\mathbb{S}) = 1$ and for the standard SECR model dim($\mathbb{S}) = 2$ (see Examples 1-3 above). The number of integrals in the marginal data likelihood is equal to the number of unique observed encounter histories plus one (corresponding to the encounter history of not being observed). Each integral can, in general, be approximated using standard integration techniques, such as Gauss-Hermite quadrature, grid-based approaches etc. Thus the computational efficiency of this approach will be dependent on dim($\mathbb{S}$) and the number of unique encounter histories observed. For closed population models, dim($\mathbb{S})$ is typically very small. This (approximate) likelihood can be estimated using standard optimisation techniques to obtain the associated MLEs of the model parameters. 

\subsubsection{Complete data likelihood}\label{comp}

The Bayesian complete data likelihood approach specifies the unobserved individual heterogeneity terms, $\bfepsilon^{Mis}$, as auxiliary variables (or additional parameters). The joint posterior distribution of the parameters and auxiliary variables is then formed and given by,
\begin{eqnarray*}
\pi(N, \bftheta, \bfeta, \bfepsilon^{Mis}|\bfx, \bfepsilon^{Obs}) & \propto & f_c(\bfx, \bfepsilon | N, \bftheta, \bfeta) p(N,\bftheta,\bfeta) \nonumber \\
& = & f_{\bfx}(\bfx | N, \bftheta,\bfepsilon) f_{\bfepsilon}(\bfepsilon|N,\bfeta) p(N,\bftheta,\bfeta), \label{eq:bay1}
\end{eqnarray*}
where $f_c(\bfx, \bfepsilon | N, \bftheta, \bfeta)$ denotes the complete data likelihood; $f_{\bfx}(\bfx|N, \bftheta,\bfepsilon)$ the conditional likelihood of the observed data (conditional on the full set of individual heterogeneity terms); $f_{\bfepsilon}(\bfepsilon|N,\bfeta)$ the individual heterogeneity component; and $p(N,\bftheta,\bfeta)$ the prior density specified on $N$, $\bftheta$ and $\bfeta$. The posterior density of only the model parameters, $\pi(N, \bftheta, \bfeta|\bfx,\bfepsilon^{Obs})$, is obtained by integrating out over the auxiliary variables, $\bfepsilon^{Mis}$. However, the integration is analytically intractable so that an MCMC approach is typically implemented, whereby we construct a Markov chain with stationary distribution equal to the joint posterior distribution, $\pi(N, \bftheta, \bfeta, \bfepsilon^{Mis}|\bfx, \bfepsilon^{Obs})$, and subsequently estimates of the marginal posterior summary statistics of interest are obtained. 

\comment{The integration is analytically intractable so that an MCMC approach is typically implemented, whereby we construct a Markov chain with stationary distribution equal to the joint posterior distribution $\pi(N, \bftheta,\bfepsilon^{Mis}|\bfx, \bfepsilon^{Obs})$. Once the chain has converged to the stationary distribution, realisations of the Markov chain can be regarded as a dependent sample from the full joint posterior distribution. Considering only the realisations of the parameters $N$ and $\bftheta$ essentially integrates out the auxiliary variables, $\bfepsilon^{Mis}$ from which Monte Carlo estimates of the quantities of interest can be obtained for the joint posterior distribution, $\pi(N,\bftheta|\bfx, \bfepsilon^{Obs})$.}

\comment{performed by obtaining a set of sampled values from the joint posterior distribution, $\pi(N, \bftheta,\bfepsilon|\bfx)$ and marginalising to obtain a set of sampled values from $\pi(N,\bftheta|\bfx)$, which are subsequently used to obtain Monte Carlo estimates of the quantities of interest. }

An additional computational model fitting difficulty arises since $\bfepsilon = \{\bfepsilon_1,\dots,\bfepsilon_N\}$ and hence $\bfepsilon$ is itself a function of the unknown parameter, $N$. To address this issue \cite{kinb08} describe a reversible jump (RJ) MCMC algorithm for $M_h$-type models that is able to explore the joint posterior distribution, where the number of parameters is able to vary within the constructed Markov chain. This involved writing bespoke computer code. Alternatively, \cite{dure05,roydl07,roykgk09,royd12} use data augmentation techniques that can be fitted in BUGS/JAGS. The underlying idea is to specify a super-population of size $M$, with associated individual random effect terms $\bfepsilon_i$ for $i=1,\dots,M$. The encounter histories for individuals $n+1,\dots,M$ correspond to not being observed within the study. Within the MCMC algorithm, the random effect term for each individual in this super-population is imputed in addition to a binary indicator variable, $z_i$ for $i=1,\dots,M$, identifying which members of the super-population are members of the target population of interest (by definition $z_i = 1$ for $i=1,\dots,n$, i.e. for all individuals observed at least once within the study). This binary indicator variable has been implemented using two different techniques each with different consequences. \cite{dure05} specify the binary variables, such that $z_1,\dots,z_N=1$ and $z_{N+1},\dots,z_M = 0$ (i.e. the indicator variables are ordered); whereas \cite{roydl07,roykgk09} do not induce any such structure on the indicator variables relating to unobserved individuals, setting $z_i=1$ for $i=1,\dots,n$ and modelling each indicator variable $z_i$ for $i=n+1,\dots,M$. The estimate of $N$ is obtained as the sum of non-zero indicator variables, i.e. $N = \sum_{i=1}^M z_i$. In other words \cite{dure05} define the indicator variables, conditional on $N$, whereas \cite{roydl07,roykgk09} define $N$, conditional on the indicator variables. For ease of reference we refer to the complete data likelihood data approach of \cite{dure05} as CD:DE (complete data: Durban and Elston) and of \cite{roydl07,roykgk09,royd12} as CD:R (complete data: Royle).

Several issues arise with regard to these super-population data augmentation approaches. For both approaches $M$ needs to be specified and corresponds to an upper bound for the total population size. This necessarily leads to a trade-off between the size specified for $M$ and the computational speed of the code. The larger the value of $M$, the greater the computational time due to the imputation of the random effect term (and binary indicator variable for CD:R) for each individual in the super-population. Too small a value for $M$ will lead to a truncation of the posterior distribution and biased inference. In addition, for CD:R, since $N$ is derived as a deterministic function of the indicator variables, it has a more limited prior specification (see Section \ref{prior} for further discussion regarding prior specification). Alternatively for the approach of CD:DE, due to the more restricted nature of the indicator variable specification, mixing issues can arise. To aid in the efficiency of the computational algorithm \cite{dure05} advocate the use of a pseudo-prior for the corresponding random effect terms for individuals not in the population (i.e. for $\epsilon_i$ for all $i=N+1,\dots,M$). The pseudo-prior is obtained from an initial MCMC run, using the estimated posterior distribution for the random effect of an unobserved individual. For further discussion of data augmentation techniques (particularly focusing on CD:R), see for example, \cite{lin13} and \cite{schb14}.

In general, without any prior information, the choice of analysis (classical marginal data likelihood or Bayesian complete data likelihood) may be data dependent. In general, for a given dataset, there is a computational trade-off between these different approaches. The marginal data likelihood requires the numerical approximation of the integrals over the individual random effects; the complete data likelihood is fast to evaluate but the individual random effects need to be updated within the MCMC algorithm (using either RJMCMC or a super-population approach). To avoid the use of explicitly approximating multiple integrals or the need to use a super-population or trans-dimensional algorithm, we propose a hybrid semi-complete data likelihood approach. This involves numerical integration for that part of the likelihood corresponding to unobserved individuals (as in the marginal likelihood approach), while for the observed individuals any unobserved individual heterogeneity terms are treated as auxiliary variables within a data augmentation approach (as in the complete data likelihood approach). In this case, the number of auxiliary variables is known so that the dimension of the parameter space is known and fixed. Standard BUGS/JAGS software readily accommodates this approach, which involves approximation of only a single integral of dimension dim($\mathbb{S}$). We describe this approach in more detail next.

\section{Semi-complete data likelihood}\label{sect3}

We propose a \emph{semi-complete} data likelihood approach, combining the complete data likelihood for the individuals that are observed within the study (i.e. individuals $i=1,\dots,n$), with a marginal data likelihood for the individuals that are not observed within the study (i.e. individuals $i=n+1,\dots,N$). The semi-complete likelihood is expressed in the form,
\begin{linenomath}
\[
f_s(\bfx, \bfepsilon_{1:n}| N, \bftheta, \bfeta) = f_{\bfx^*}(\bfx| N, \bftheta, \bfeta, \bfepsilon_{1:n})f_{\bfepsilon}(\bfepsilon_{1:n} |  N, \bfeta)
\]
\end{linenomath}
where $\bfepsilon_{1:n} = \{\bfepsilon_1,\dots,\bfepsilon_n\}$; $f_{\bfx^*}(\bfx| N, \bftheta, \bfeta, \bfepsilon_{1:n})$ denotes the conditional likelihood of the capture histories conditional on the model parameters ($N$, $\bftheta$ and $\bfeta$) and individual heterogeneity terms for the observed individuals only ($\bfepsilon_{1:n}$); and $f_{\bfepsilon}(\bfepsilon_{1:n} | N, \bfeta)$ the joint probability density function of the individual heterogeneity component for the observed individuals. Further, we have the following conditional likelihood functions: $f_{\bfx^*}(\bfx_{1:n} | N, \bftheta, \bfepsilon_{1:n})$ for the capture histories of the observed individuals only, conditional on the model parameters and individual heterogeneity terms for the observed individuals (dropping the dependence on $\bfeta$ since these are conditionally independent given $\bfepsilon_{1:n}$); $f_{\bfx^*}(\bfx_{n+1:N} | N, \bftheta, \bfeta)$ for the capture histories of the unobserved individuals, conditional on the model parameters; and $f_{x^*}(\bfx_i | \bftheta, \bfeta) $ for the capture history for unobserved individual $i=n+1,\dots,N$, given the capture probability and individual heterogeneity model parameters (in the latter two cases dropping the conditioning on $\bfepsilon_{1:n}$). Then, letting $\bfx_{a:b} = \{\bfx_a,\dots,\bfx_b\}$, we can express the conditional likelihood in the form:
\begin{eqnarray}
f_{\bfx^*}(\bfx| N, \bftheta, \bfeta, \bfepsilon_{1:n}) & = & f_{\bfx^*}(\bfx_{1:n} | N, \bftheta, \bfepsilon_{1:n}) f_{\bfx^*}(\bfx_{n+1:N} | N, \bftheta, \bfeta) \nonumber \\
& \propto & \frac{N!}{(N-n)!}  \prod_{i=1}^n f_x(\bfx_i | \bftheta, \bfepsilon_i) \times \prod_{i=n+1}^N f_{x^*}(\bfx_i | \bftheta) \nonumber \\
& = & \prod_{i=1}^n f_x(\bfx_i | \bftheta, \bfepsilon_i) \times \frac{N!}{(N-n)!} \left( 1-p^* \right)^{N-n}, \label{eq:semi}
\end{eqnarray}
where $1-p^*$ denotes the probability of not being observed within the study (or conversely $p^*$ denotes the probability of being seen at least once within the study) such that,
\begin{linenomath}
\begin{equation}\label{int}
1-p^* = \int_{\bfepsilon_{\bftomega} \in \mathbb{S}} f_x(\bfomega = \bfzero | \bftheta, \bfepsilon_{\bfsomega}) f_\epsilon(\bfepsilon_{\bfsomega}|\bfeta)d\bfepsilon_{\bfsomega},
\end{equation}
\end{linenomath}
and $\bfomega = \bfzero$ denotes the encounter history of a single individual who is unobserved within the study; $f_x(\bfomega = \bfzero | \bftheta, \bfeta, \bfepsilon_{\bfsomega})$ the conditional likelihood function associated with an individual not observed within the study and $f_\epsilon(\bfepsilon_{\bfsomega}|\bftheta,\bfeta)$ the probability density function of the associated individual heterogeneity terms for an individual not observed within the study. The product in equation (\ref{eq:semi}) corresponds to the likelihood of the encounter histories, for an individual observed at least once within the study, conditional on the individual heterogeneity terms. The latter term corresponds to the contribution to the likelihood relating to the unobserved individuals.

An alternative (equivalent) model specification is given by
\begin{eqnarray}
f_{\bfx^*}(\bfx| N, \bftheta,\bfeta,\bfepsilon_{1:n}) & \propto & \frac{1}{(p^*)^n} \prod_{i=1}^n f_x(\bfx_i | \bftheta, \bfepsilon_i) \times \frac{N!}{(N-n)!} (p^*)^n (1-p^*)^{N-n}, \label{eq:semi1}
\end{eqnarray}
where $p^*$ is as above. The first term corresponds to the conditional likelihood of the observed capture histories, \emph{given} that each of these individuals has been observed within the study and the corresponding individual heterogeneity terms. The second term corresponds to the Binomial probability of observing the number of individuals in the study, \emph{given} the total population size.

We note that the semi-complete likelihood reduces to a single integral (over the dimension of the individual heterogeneity terms, i.e. dim($\mathbb{S}$)). This is in contrast to the marginal data likelihood which is a product of integrals (see Section \ref{obs}), where the number of additional integrals corresponds to the number of unique encounter histories observed. 

\subsection{Bayesian implementation}

Notationally, we let $\bfepsilon_{1:n}^{Obs}$ and $\bfepsilon_{1:n}^{Mis}$ denote the set of observed and unobserved individual heterogeneity terms for the observed individuals, respectively. The joint posterior distribution for the model parameters and unobserved individual heterogeneity terms for the observed individuals is given by,
\begin{eqnarray}
\pi(N,\bftheta,\bfeta,\bfepsilon_{1:n}^{Mis}|\bfx, \bfepsilon_{1:n}^{Obs}) & \propto & f_s(\bfx, \bfepsilon_{1:n} | N, \bftheta, \bfeta) p(N,\bftheta, \bfeta) \nonumber \\
& = & f_{\bfx^*}(\bfx |N, \bftheta,\bfeta,\bfepsilon_{1:n}) f_{\bfepsilon}(\bfepsilon_{1:n}|N,\bfeta) p(N,\bftheta,\bfeta),\label{eq:bay2}
\end{eqnarray}
where $f_s(\bfx, \bfepsilon_{1:n} | N,\bftheta,\bfeta)$ is the semi-complete data likelihood. Note that, as is typically the case, we assume that the priors specified for the total population size and model parameters are independent, so that $p(N,\bftheta,\bfeta) = p(N)p(\bftheta)p(\bfeta)$.

We use a standard Bayesian data augmentation approach for obtaining inference on the posterior distribution of interest, $\pi(N,\bftheta,\bfeta|\bfx, \bfepsilon^{Obs}_{1:n})$. The number of auxiliary variables needed within this Bayesian data augmentation approach, using the semi-complete likelihood, is fixed and simply equal to $|\bfepsilon^{Mis}_{1:n}|$ (i.e. the auxiliary variables correspond to the number of unobserved individual heterogeneity terms of observed individuals). This is in contrast to the use of the joint posterior distribution of the model parameters and all unobserved individual heterogeneity terms, $\bfepsilon^{Mis}$, given in equation (\ref{eq:bay1})), since $\bfepsilon^{Mis} = \{\bfepsilon^{Mis}_{1:n},\bfepsilon_{n+1:N}\}$ where $N$ is a parameter to be estimated. A number of different approaches have been proposed to fit individual heterogeneity models. These include trans-dimensional algorithms using reversible jump MCMC \citep{kinb08}, a joint posterior conditional MCMC algorithm \citep{fie99} for Rasch-type ($M_{th}$) models and super-population data augmentation techniques. The first two approaches require bespoke code, while the super-population data augmentation approaches can be implemented within BUGS/JAGS \citep{dure05,roydl07,royd12} but require the specification of an upper bound $M$ and imputation of the $(M-n)$ individual random effect terms $\bfepsilon_{n+1:M}$ (and dependent on the exact coding approach, $M$ binary indicator variables). 

Using the semi-complete data likelihood and corresponding posterior distribution given in equation (\ref{eq:bay2}), including only the heterogeneity terms for the observed individuals, permits standard (non-trans-dimensional) MCMC updating algorithms (such as the Metropolis-Hastings algorithm) to obtain inference on the parameters $\bftheta$, $\bfeta$ and $N$. However, the semi-complete data likelihood removes the necessity to impute the terms $\bfepsilon_{n+1:M}$ and the need to specify an upper bound on the total population size, in general (see Section \ref{prior}). Consequently, the models can be immediately fitted within BUGS/JAGS packages (see Section \ref{sect4} for further discussion and the Web Appendix for example JAGS code), with an explicit prior distribution specified on $N$. The trade-off of using the posterior distribution with semi-complete data likelihood, given in equation (\ref{eq:bay2}), is that the integral in equation (\ref{int}) needs to be explicitly (numerically) estimated. However, in general this will be of very low dimension (often only one or two dimensions) for closed population models and so computationally fast and able to be accurately estimated (for example using Gaussian quadrature). We compare the complete data likelihood and semi-complete data likelihood approaches in Section \ref{sect4} using JAGS for two different applications. 

\subsection{Prior specification for $N$}\label{prior}

We briefly discuss possible prior distributions that are commonly specified on $N$ and the corresponding Bayesian (and BUGS/JAGS) implementation. For the Bayesian data augmentation approach of \cite{roydl07} (approach CD:R), the prior on $N$ is only defined implicitly, given the prior specification on the indicator function relating to the probability that an individual in the super-population is a member of the population of interest, denoted $\psi$. The most common form of induced prior on $N$ is the Uniform prior. However, \cite{lin13} showed that the uninformative prior $\psi \sim U[0,1]$ which induces the discrete uniform prior on $N$ can lead to undesirable properties. \cite{lin13} therefore recommended the prior $\psi \sim Beta(0.001,1)$ which is easy implemented in BUGS/JAGS and induces an approximate Jeffreys' prior on $N$. More generally, specifying the prior $\psi \sim Beta(a,b)$ induces the prior $N \sim Beta-Binomial(M,a,b)$, where $M$ is the super-population upper bound. This is a fairly flexible prior structure, but the computational limitations with regard to specifying a suitable value of $M$ remain. 

For the complete data likelihood approach of \cite{dure05} (approach CD:DE) and the semi-complete data likelihood approach an explicit prior is directly specified on $N$. Thus, any arbitrary distribution (specified on the set of non-negative integers) can be specified on the total population size. For example, Jeffreys' prior is a commonly used uninformative prior, given by $p(N) \propto N^{-1}$ (see for example, \cite{mady97,kinb08}). We note that specifying Jeffreys' prior, and using the semi-complete data likelihood expression given in equation (\ref{eq:semi1}) leads to a standard posterior conditional distribution for $N$, i.e.,
\begin{linenomath}
\[
(N-n)|\bfx,\bftheta,\bfeta \sim Neg-Bin(n,p^*),
\]
\end{linenomath}
for $p^*$ given in equation (\ref{eq:semi})\footnote{We use the form of the Negative Binomial distribution such that for $X \sim Neg-Bin(n,q)$ the probability mass function is given by,
\[
f(x) = \frac{(x+n-1)!}{x!(n-1)!} q^n (1-q)^{x}.
\]
This is the functional form of the distribution used with BUGS/JAGS.}. Consequently, for Jeffreys' prior, the Gibbs sampler can be implemented for updating $N$ within the MCMC algorithm. In general, if the prior or posterior conditional distribution for $N$ is of (closed or) standard form this also simplifies the specification of the model in BUGS/JAGS, since this prior or posterior conditional distribution can be explicitly specified in the model component (see the Web Appendix for sample JAGS code for the above Negative-Binomial posterior conditional distribution case). See also \cite{fie99} for further discussion.

Alternative prior distributions include $p(N) \propto N^{-c}$ for some positive constant $c$, proposed by \cite{fie99}. For $c > 1$ the tail of the distribution for $N$ decays faster than for Jeffreys' prior; while $c < 1$ leads to a heavier tailed distribution. Alternatively, for an informative prior distribution for $N$, a Poisson or Poisson-Gamma (equivalently a Negative-Binomial) prior distribution is often specified on $N$ \citep{kinb01}. It can also be noted that specifying $N \sim Po(\lambda)$ and $\lambda \sim \Gamma(\delta, \delta)$ for small $\delta$ provides another approximate Jeffreys' prior for $N$. These alternative prior distributions are able to be implemented within BUGS/JAGS (typically using the zeros or ones trick, \citet{bugs13} - see the Web Appendix for associated sample JAGS code). 

\subsection{Special case}\label{special}

We note that the approach presented by \cite{bons14} is a special case of the semi-complete data likelihood approach applied to a covariate model. In particular, \cite{bons14} consider a time invariant individual covariate model given in Example 1 of Section \ref{sect2}. Using the terminology presented above, so that the notation differs to that given in \cite{bons14}, they describe the particular case where $\bfepsilon^{Obs} = \bfepsilon_{1:n}$ and $\bfepsilon^{Mis} = \bfepsilon_{n+1:N}$. In other words, the individual heterogeneity terms are known for individuals observed within the study (though it is implied in their discussion that the approach is more generally applicable). The posterior distribution is then formed analogous to Equation (\ref{eq:bay2}). The probability of not being observed within the study, given in Equation (\ref{int}) is estimated using Monte Carlo integration. 
\comment{
\begin{enumerate}
\item For closed population model:
\begin{itemize}
\item 
\begin{eqnarray*}
f(\bfx| N, \bftheta,\epsilon_1,\dots,\epsilon_n,\sigma^2) & \propto & \frac{1}{(p^*)^n}\prod_{i=1}^n \prod_{t=1}^T p_{it}^{x_{it}} (1-p_{it})^{1-x_{it}} \\
& \times & \frac{N!}{n!(N-n)!} (p^*)^n (1-p^*)^{N-n},
\end{eqnarray*}
where,
\[
1-p^* = \int_{\epsilon \in \mathbb{R}} \prod_{t=1}^T [1-f^{-1}(\alpha+\beta_t+\epsilon)] \frac{1}{\sqrt{2 \pi \sigma^2}} \exp\left(-\frac{\epsilon^2}{2 \sigma^2} \right) d\epsilon,
\]
corresponding to the probability of not being observed within the study.
\item This model can again be fitted in OpenBUGS (see file ``bugsint2.odc'' for model $M_h$).
\item (Note that numerical problems can arise when the probability of being observed within the study is close to the boundary).
\end{itemize}
\item For SECR models:
\begin{itemize}
\item \textbf{To complete}
\end{itemize}
\end{enumerate}}

\section{Examples}\label{sect4}

We consider two real examples, relating to model $M_h$ and SECR, described in Section \ref{sect2}. We note that as with all performance metrics for comparing the efficiency of different model-fitting approaches these are dependent on numerous factors, such as the programming language, specific application, data, model specification (including the pseudo-priors specified for the super-population approach), initial starting values and machine used. In order to be able to draw sensible comparisons for each example we present results obtained from same machine and language using the JAGS codes provided in Web Appendix A. 

\subsection{Model $M_h$ - snowshoe hares}\label{hares}

To demonstrate our proposed semi-complete data likelihood approaches for model $M_h$, we revisit the snowshoe hare data originally examined in the seminal paper of \cite{otis78} and subsequently analyzed by many others \cite[for example][]{coua99,dorr03,roydl07,lin13}.  Over $T=6$ days of trapping, $n=68$ hares were captured with observed frequencies $\bfn = (25,22,13,5,1,2)^\prime$ where $n_t=\sum_{i=1}^{n} I(y_i=t)$ and $y_i=\sum_{j=1}^T x_{ij}$ for $t=1,\dots,T$.  We assume logit$(p_{it}) = \alpha+\epsilon_i$  and $\epsilon_i \sim N(0,\sigma^2)$ for $i=1,\ldots,N$ and $t=1,\ldots,T$, with $\bftheta = \{ \alpha\}$ and $\bfeta = \{\sigma^2\}$.  

We fit the semi-complete data likelihood and complete data likelihood Bayesian super-population (CD:R and CD:DE) approaches in R \citep{RTeam2014} using the \texttt{rjags} package \cite[][see Web Appendix A for JAGS code]{Plummer2013}. 
For each analysis we specify the priors, $\alpha \sim N(0,100)$ and $\sigma^2 \sim \Gamma^{-1}(0.01, 0.01)$. We specify Jeffreys' prior for $N$, for the semi-complete data likelihood and CD:DE.  For ease of comparison with CD:R we set $\psi \sim Beta(0.001,1)$, which induces an approximate (truncated) Jeffreys' prior for $N$ on ${1,\ldots,M}$ \citep{lin13}. We note that we consider two JAGS specifications for the semi-complete data likelihood. The first approach (SCD1) uses the Jeffreys' prior specification for $N$ explicitly in the model component of the code. However, since Jeffreys' prior is improper we need to specify an upper bound for $N$, which we again denote by $M$ (essentially this is a truncated Jeffreys' prior at $M$).  The second approach (SCD2) specifies the (predictive) posterior conditional distribution for $N-n$, which is of Negative-Binomial form (see Section \ref{prior}). 

Following \cite{lin13}, we specify an upper bound of $M=1000$ for the maximum total population size for the complete data likelihood super-population approaches and the first semi-complete data likelihood approach (SCD1) in JAGS. For the semi-complete data likelihood approach, the integral in Equation (\ref{int}) is evaluated using Gauss-Hermite quadrature:
\begin{eqnarray}
1-p^* \approx \sum_{j=1}^q \frac{w_j}{ \sqrt{\pi} \left[ 1+ \exp{\left(\sqrt{2} \sigma v_j+\alpha \right)} \right]^T },
\label{eq:ghquad}
\end{eqnarray}
where $w_j$ and $v_j$ are the weights and nodes corresponding to $q$ quadrature points \cite[sensu][]{McClintockEtAl2009}. The degree of accuracy of this approximation increases with $q$, and larger $q$ is required for larger $\sigma$.  For our analyses, we specify $q=100$.

For each approach, we ran three chains of 10 million iterations (after initial pilot tuning and burn-in) from overdispersed starting values, thinning the realisations by 10 for memory storage purposes.  Chain convergence was assessed based on visual inspection and Brooks-Gelman-Rubin diagnostics (no lack of convergence was identified).  On a computer running 64-bit Windows 7 (3.4GHz Intel Core i7 processor, 16Gb RAM), the analyses required about 6.1 hrs for the first semi-complete data likelihood (prior distribution for $N$ specified) approach, 6.0 hrs for the second semi-complete data likelihood (posterior conditional distribution for $N-n$ specified), 35.1 hrs for CD:R and 83.3 hours for CD:DE. We note that the run times should be interpreted comparatively, as they will in general differ across different computers as a result of different processors, operating systems etc. The marginal posterior summaries are provided in Table \ref{tab:snowshoehare}, coupled with the effective sample sizes (per second) for each approach.

\begin{table}
\caption{Posterior summaries for snowshoe hare abundance $(N)$ under model $M_h$ using the semi-complete data likelihood (SCD) approach, CD:R and CD:DE. The semi-complete data likelihood approaches correspond to specifying the prior for $N$ (SCD1) and the posterior conditional distribution for $N-n$ (SCD2) in the model component of the JAGS code. For SCD1, CD:R and CD:DE, we specify an upper limit of $M=1000$.  Effective sample size (ESS) and effective sample size per second (ESS/s) are included for each approach. A total of 30 million iterations are used in each case with the realisations thinned by 10.}
\begin{tabular}{c c c c c c c c}
& & & & & & & \\
method & mean & median & SD & 95\% CI  & ESS & ESS/s \\
\hline
SCD1	&	100.3	&	93	&	32.8	&	(74, 171)	& 168347	& 7.67	\\
SCD2	&	101.1	&	93	&	74.9	&	(74, 173)	& 167680	& 7.74	\\
CD:R	&	100.6	&	93	&	32.7	&	(74, 171)	& 13080		& 0.10 \\
CD:DE	&	101.3	&	93	&	36.2	&	(74, 178)	&  9626		& 0.03 \\
\end{tabular}
\label{tab:snowshoehare}
\end{table}

Although setting $M=1000$ may appear conservative, this did appear to influence the skewness of the right tail of the marginal posterior distribution for $N$ relative to the (unbounded) posterior distribution for $N$ when using the second semi-complete data likelihood approach (SCD2).  We therefore reran the first semi-complete data likelihood (SCD1) analysis with $M=10000$ leading to posterior summary results more similar to the second complete data likelihood approach ($N$ posterior $\mbox{mean}=100.9$, $\mbox{median}=93$, $\mbox{SD}=56.1$, $95\%$ credible interval (CI) = (74, 172)), but with noticeably reduced effective sample size $(\mbox{ESS}=74928)$ and increased computation time $(\mbox{ESS/s}=2.81)$.  Nevertheless, specifying larger $M$ for the first semi-complete data likelihood approach comes at considerably less computational cost compared to the super-population complete data likelihood approaches (CD:R and CD:DE). Avoidance of the need to specify $M$ when using BUGS/JAGS remains an advantage of the general semi-complete data likelihood approach (this is true even when using Jeffreys' prior on $N$ by specifying the posterior conditional distribution for $N-n$ in the model component of the code). 

For approach SCD1, using an explicit Negative-Binomial or Beta-Binomial approximation to Jeffreys' prior (code provided in Web Appendix A) unsurprisingly lead to similar results in terms of ESS and ESS/s as for the use of the explicit (truncated) Jeffreys' prior. However, within the model specification code, using the distributions' hierarchical form where an auxiliary variable is introduced for the Poisson mean or Binomial probability and imputed within the MCMC algorithm lead to lower ESS and ESS/s as a result of poorer mixing due to posterior correlation between parameters. We do not consider these prior specifications further. 

Finally, we note that $q=100$ appeared to be sufficient in the Gauss-Hermite quadrature approach for these analyses, but in general proper specification of $q$ will be case dependent.  For example, using our estimated posterior median $\alpha=-1.2$ and the 99.9\% quantile $\sigma=3.3$, Equation (\ref{eq:ghquad}) with $q=100$ is accurate to a precision of five decimal places.  However, for $\sigma=10$, $q=100$ it is only accurate to two decimal places.  Care must therefore be taken when specifying $q$ using the semi-complete data likelihood approach in JAGS. If computation speed is of little concern Equation (\ref{int}) could alternatively be approximated in OpenBUGS using the inbuilt \texttt{integral} function, which also has an inbuilt default value for $q$.

\subsection{Model SECR - gibbons}\label{gibbons}

To illustrate the proposed semi-complete data likelihood approach in the context of SECR models we use acoustic survey data from a population of northern yellow-cheeked gibbon from northeastern Cambodia. These data were collected from 13 replicate survey locations, each consisting of a 3 by 1 linear array of listening posts spaced 0.5km apart. Each listening post was manned by a single human observer who recorded the timing of calls at each and an estimated compass bearing to each detected gibbon group. Recaptured groups were determined using the estimated bearings and detection times. Over $T=1$ survey days a total of $n=77$ gibbon groups were detected across the 13 arrays. We specify the half-normal function for $g$ of the form,
\begin{linenomath}
\[
p_{ijt} = \exp\left(-\frac{||\bfu_j-\bfeps_i||^2}{2 \sigma^2}\right).
\]
\end{linenomath}
For each analysis we specify the prior $\sigma \sim U[0,10]$ and assume that the home range centres are uniformly distributed over the given area, i.e. $f_\epsilon(\bfepsilon_i | \bfeta) = \frac{1}{A}$ where $A$ is the area of $\mathbb{S}$ for each $i=1,\dots,N$ (in this case $A = 546$km$^2$). Thus we set $\psi \sim Beta(0.001,1)$ for the super-population approach CD:R and Jeffreys' prior for $N$ for the complete data likelihood (CD:DE) semi-complete likelihood approaches. 

As in Section \ref{hares} we fit both forms of the semi-complete data likelihood (Equations (\ref{eq:semi}) and (\ref{eq:semi1})) and the super-population complete data likelihood Bayesian approaches CD:R and CD:DE using the \texttt{rjags} package \cite[][see Web Appendix B for sample JAGS code]{Plummer2013}. For the complete data likelihood approaches and first semi-complete data likelihood (specifying Jeffreys' prior on $N$ within the model component of the JAGS code) we specify an upper bound of $M=1000$ for the discrete support of $N$. For both semi-complete likelihoods the integral in Equation (\ref{int}) was approximated by a summation over a rectangular grid of 4200 points. Note that a suitable choice of grid will be case dependent, with increases in accuracy resulting from greater spatial extents and decreased distances between neighbouring grid points, but at the expense of computational time. An exploratory analysis suggested that the grid used was relatively conservative, achieving good numerical accuracy. 

To compare the performance of the different approaches, each MCMC algorithm is run for 500,000 iterations, following a burn-in period of 10,000 iterations (no lack of convergence was identified for simulations of this length). On a computer running Windows Server 2008 R2 Enterprise (3.1GHz Intel Xeon CPU E5-2687, 256Gb RAM), the analyses required about 46.6 minutes for the first semi-complete data likelihood (SCD1; specifying (truncated) Jeffreys' prior on $N$ in the model component) approach, 42.3 minutes for the second semi-complete data likelihood (SCD2; specifying the posterior conditional distribution for $N-n$), 2.5 hours for CD:R and 6.8 hours for CD:DE. As for the snowshoe hare example, marginal posterior summaries were similar for all parameters using all approaches, but the semi-complete data likelihood approaches required far less computation time and yielded greater effective sample sizes than the data-augmented complete data likelihood approaches (Table \ref{tab:gibbons}).

\begin{table}
\caption{Posterior summaries for gibbon group abundance $(N)$ under the SECR models using the semi-complete data likelihood (SCD) approach, CD:R and CD:DE. The semi-complete data likelihood approaches correspond to specifying the prior for $N$ (SCD1) and the posterior conditional distribution for $N-n$ (SCD2) in the model component of the JAGS code. For SCD1, CD:R and CD:DE, we specify an upper limit of $M=1000$. Effective sample size (ESS) and effective sample size per second (ESS/s) are included for each approach. A total of 500,000 iterations are used in each case.}
\begin{tabular}{c c c c c c c c}
& & & & & & & \\
model & mean  & median & SD    & 95\% CI    & ESS  & ESS/s \\
\hline
SCD1  & 357.1 & 328    & 176.2 & (119, 766) & 2763 & 1.01  \\
SCD2  & 357.7 & 327    & 178.4 & (120, 775) & 3872 & 1.56  \\
CD:R   & 355.3 & 326    & 176.9 & (118, 768) & 865  & 0.09 \\
CD:DE & 362.7 & 338 & 173.2 & (122, 765) & 622 & 0.03
\end{tabular}
\label{tab:gibbons}
\end{table}

\section{Discussion}\label{sect5}

\comment{
\begin{itemize}
\item Pros/cons of different approaches:
\begin{itemize}
\item Prior specification on $N$;
\item Computation for data augmentation approach depends on number of unobserved individuals ($N-n$);
\item Issues with super-population limit specification (Link - Ecology).
\item BUGS codes written can be extended to model $M_{tbh}$ - however there are syntax limitations for large numbers of capture events (are there ways around these?) and range of SECR models (multiple-individual physical traps; motion sensors permitting multiple sightings)
\item Alternative random effect models can be fitted using analogous approach (e.g. $t$-distribution etc.).
\item \texttt{integrate} function in OpenBUGS for 1-D integration. 
\end{itemize}
\item Note - Bonner and Schofield (2013) - Monte Carlo in MCMC approach with application to capture-recapture data using single-continuous individual covariate. Their approach is not readily implemented in BUGS/JAGS; this approach can be viewed as a special case that is easily implemented in BUGS/JAGS (they mention numerical quadrature could be used for a scalar covariate - but do not actually do this). - Intro!
\end{itemize}}

For closed population models, the semi-complete data likelihood specifies the joint probability density function of the model parameters and associated unobserved individual heterogeneity terms for only those individuals observed, conditional on the observed capture histories and observed individual heterogeneity components. This likelihood is specified as an integral of the individual heterogeneity component for the unobserved individuals. The integral is analytically intractable but of dimension equal to the dimension of the individual heterogeneity component of the model, and hence typically small. This permits the the use of standard (efficient) numerical approximation techniques to estimate the integral (for example, in OpenBUGS, the inbuilt \texttt{integral} function can be used to conduct one dimensional integration; with similar inbuilt functions in \texttt{R} for one or multi-dimensional integrals). The semi-complete data likelihood approach can be applied to a range of different individual heterogeneity models.

Using this semi-complete data likelihood within a Bayesian analysis of closed capture-recapture data in the presence of individual heterogeneity, removes the need for trans-dimensional algorithms to explore the posterior distribution of the parameters due to the ``unknown number of parameters'' problem. Consequently, the models can be fitted efficiently in standard software, such as BUGS/JAGS without using a super-population approach. The semi-complete data likelihood approach is significantly more efficient than the previous super-population approaches, as demonstrated in Section \ref{sect4}, where the improvement for the examples that we considered using the codes provided in the Web Appendix was up to two orders of magnitude. The improvement is in terms of both computational time and effective sample sizes (as a result of improved mixing within the MCMC algorithm). The efficiency of the super-population approaches is heavily dependent on the upper limit specified for the super-population, $M$. This makes the Bayesian approach feasible for fitting to a significantly wider range of data, particularly for spatially explicit capture-recapture, where the use of a Bayesian data augmentation technique can be particularly inefficient. In general, the ESS and ESS/s for the different approaches is dependent on numerous factors including the exact form of the model specification, the pseudo-priors specified in the super-population approach, initial starting values and computer on which the simulations are being run. 

This semi-complete data approach has been developed for closed population models in the presence of individual heterogeneity. As discussed in Example 1 of Section \ref{sect2} the inclusion of additional observable individual level covariates is immediate and can be seen to be a generalisation of the Monte Carlo in MCMC approach proposed by \cite{bons14} (see Section \ref{special}). The individual heterogeneity terms correspond to the covariate values and are typically known when individuals are observed, though this need not be the case (missing covariate values for individuals observed within the study can again be treated as auxiliary variables within the complete data likelihood component). In the presence of time-varying continuous individual covariates the increase in dimension of the necessary integral in the associated marginal data likelihood can be reduced by efficiently approximating the underlying state process as a hidden Markov model \citep{lank13}. The approach can also be immediately applied to other forms of data. For example, these include stopover models permitting arrivals to, and departures from, the study population \citep{ple09} and conventional distance sampling \citep{buc01}. For the latter case the capture history is a univariate binary term (1 if an individual is observed and 0 if unobserved), the individual heterogeneity component is the perpendicular distance of the individual from the line/point transect (known for observed individuals), assumed to have a uniform distribution (for line transects) or triangular distribution (for point transects), see for example, Equation (7.10) on page 141 of \cite{bor02}. Further work lies in identifying and developing similar approaches for different forms of data. In addition, for more general Bayesian analyses, highly correlated parameters often leads to inefficient MCMC algorithm, due to poor mixing. To address this issue, a reparameterisation may often be used and/or block-updates implemented. An alternative approach, motivated by this semi-complete data approach, would be to identify and integrate out (using a numerical approximation) the highly correlated parameters. This is an area of current research. 

\section*{Acknowledgments}

The authors would like to thank two referees and an AE for helpful comments on an earlier version of the manuscript. We would also like to thank Dr Benjamin Rawson (Conservation International, Cambodia) for permission to use the gibbon data analysed in Section \ref{gibbons} and Dr John Durban and Prof David Elston for providing sample BUGS code for the CD:DE approach. This work was part-funded by EPSRC grant EP/I000917/1. The findings and conclusions in the paper are those of the author(s) and do not necessarily represent the views of the National Marine Fisheries Service, NOAA. Any use of trade, product or firm names does not imply an endorsement by the US Government.

\section*{Supplementary Material}
Web Appendices (referenced in Sections \ref{prior}, \ref{hares} and \ref{gibbons}) are available as supplementary material on-line.

\label{lastpage}

\end{document}